\documentclass[aps,prb,twocolumn,showpacs,superscriptaddress,nofootinbib,floatfix]{revtex4-1}

\usepackage{color}
\usepackage{amsmath}
\usepackage{amssymb}
\usepackage{graphicx}
\usepackage{bm}

\begin{document}

\title{Ab initio transport calculations: from normal to superconducting current}

\author{H. Ness}
\affiliation{Department of Physics,
	%Faculty of Natural and Mathematical Sciences,
	King's College London, Strand, London WC2R 2LS, UK}

\author{M. van Schilfgaarde}
\affiliation{Department of Physics,
	%Faculty of Natural and Mathematical Sciences,
	King's College London, Strand, London WC2R 2LS, UK}
\affiliation{National Renewable Energy Laboratory, Golden, Colorado 80401, USA}

\begin{abstract}

Applying the Bogoliubov-de Gennes equations with density-functional theory, it is possible to formulate first-principles
description of current-phase relationships in superconducting/normal (magnetic)/superconducting trilayers.  Such
structures are the basis for the superconducting analog of Magnetoresistive random access memory devices (JMRAM).
In a recent paper [\onlinecite{Ness:2022}] we presented results from the first attempt to formulate such a theory,
applied to the Nb/Ni/Nb trilayers.  In the present work we provide computational details, explaining how to construct
key ingredient (scattering matrices $S_N$) in a framework of linear muffin-tin orbitals (LMTO).

\end{abstract}

%\pacs{71.15.-m, 72.10.-d, 74.50.+r, 74.45.+c}

\maketitle

\section{Introduction}
\label{sec:intro}

In a recent paper [\onlinecite{Ness:2022}], we have combined density functional theory and the Bogoliubov-de Gennes equations 
to form a first-principles approach to the study of transport in magnetic Josephson junctions (MJJ). This method allowed us 
to predict and explain the properties of realistic MJJs such as the period of oscillation and decay of the critical current oscillations with 
the ferromagnet thickness. 
We applied our methodology to study realistic material stacks of the Nb/Ni/Nb trilayer and established that suppression of supercurrent 
is an intrinsic property of the junctions, even in absence of disorder.

To determine the supercurrent in a superconductor–normal metal–superconductor (S/N/S) junction from a density-functional approach 
(which is inherently single-particle), one needs to ``decompose'' the entire scattering process into different steps.
First we use the Andreev approximation to account for electron–hole scattering processes \cite{Beenakker:1992}, the (spin-resolved) 
Andreev reﬂection at each left S/N and right N/S interfaces is described by a reflection matrix.
Second, we assume that the main contribution to the supercurrent comes from Andreev bound states localized in the junction
(in the short junction limit) in the energy window corresponding to the superconducting gap.
The energy spectrum of the Andreev bound states can be obtained by solving an equation
\cite{Beenakker:1992,Beenakker:2004,vanHeck:2014,deJong:1995} which basically states the conservation of incoming/outgoing particle fluxes
following scattering in the normal state region and Andreev reflections at the left S/N and right N/S interfaces.
This equation involves the Andreev reflection matrices (at the left S/N and right N/S interfaces) as well as the scattering
matrices $S_N$ for electron- and $S_N^*$ hole-like waves in the central normal N region.
These single-particle (normal state) $S_N$ scattering matrices are needed the objects we obtain from density-functional theory.

In the present paper, we show in detail how to obtain such scattering matrices from the Questaal suite, which is an open
access electronic structure code based on the LMTO technique \cite{Questaal:html,Pashov:2020}.

\section{Transport in the normal state}
\label{sec:transp}

The Questaal package \cite{Questaal:html,Pashov:2020} calculates the single-particle electronic structure and includes
some many-body effect corrections as well.  It is based on the LMTO technique.

\begin{figure}
	\centering
	\includegraphics[scale=0.35]{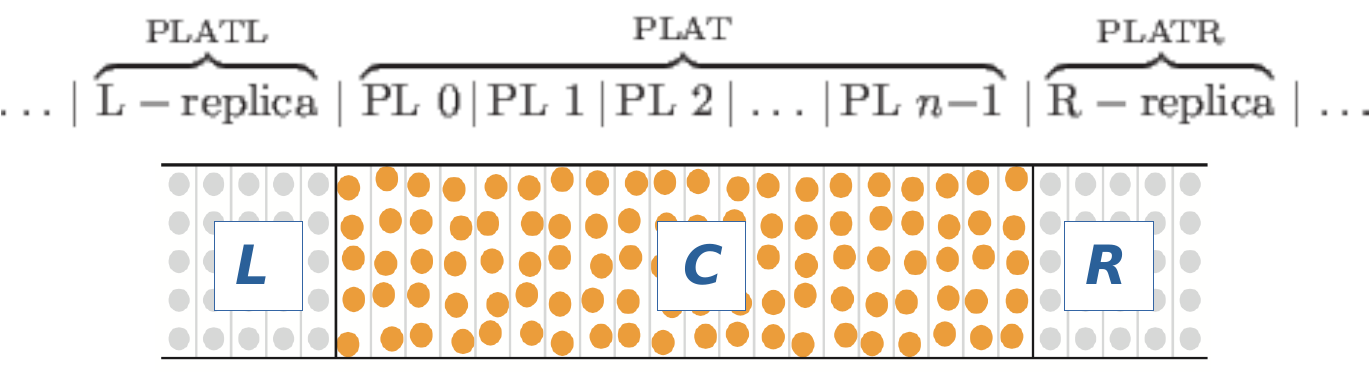}
	\caption{Schematic representation of the stacking of principal-layers PLs
	for the calculation of the transport in two-terminal $L$-$C$-$R$ junctions.}
	\label{fig:PLs}
\end{figure}

One of Questaal's packages provides the ability to calculates the full non-equilibrium (NE) transport properties of an
infinite $L$-$C$-$R$ system representing a central region $C$ cladded by two semi-infinite $L$ and $R$ leads
\cite{Faleev:2005}.  The transport properties are obtained by using the NE Green's functions (GF) formalism
\cite{Meir:1992}.
The transport calculations are done with a LMTO basis-set in the so-called Atomic Sphere Approximation. Owing to the
finite range of the basis set, the $L$-$C$-$R$ system can be described in terms of an infinite stack of principal-layers
(PLs) which interact only with their nearest neighboring PLs (Fig.~\ref{fig:PLs}).  The direction of the electronic
current is perpendicular to the PLs, and periodic boundary conditions are used within each PL.  The discretization of
the surface Brillouin zone of the corresponding PL introduces a set of transverse momentum $k_\parallel$.  Before
transport calculations are performed, the density of the $L$-$C$-$R$ must be computed self-consistently in a standard
DFT framework.  Apart from a constant shift, potentials in the semi-infinite $L$- and $R$- layers are kept frozen at the
potential of their respective bulk systems.  A dipole must form across the $C$ region so that the $L$- and $R$- Fermi
levels align.  This is generated in the course of the self-consistent cycle, and also determines the shift needed to
align the $L$- and $R$- Fermi levels.
The first account of this method was presented in Ref.[\onlinecite{MvS:1998}], and the formalism is described in detail
in Ref.[\onlinecite{Faleev:2005}], including its implementation for the non-equilibrium cases.
Self-consistency can be obtained in the non-equilibrium case, though it is not important here.

The key quantity is the total transmission probability $T(E,V)$ from which
the non-linear current can be obtained with the conventional Landauer-like
elastic scattering framework.
The linear conductance regime is simply described by $T(E_F,V=0)$ taken at
the Fermi energy $E_F$ of the $L$-$C$-$R$ system at equilibrium.
The total transmission probability $T(E,V)$ is given by
\begin{equation}
T(E,V) = \sum_{k_\parallel,\sigma} w_{k_\parallel} T(E,V;k_\parallel,\sigma) \ ,
\label{eq:ToEtot}
\end{equation}
where $w_{k_\parallel}$ is the weight associated with the transverse momentum
$k_\parallel$, and $\sigma$ is the spin of the electron.

The partial transmission probability $T(E,V;k_\parallel,\sigma)$ are obtained
from NEGF \cite{Meir:1992} as follows:
\begin{equation}
\begin{split}
& T(E,V;k_\parallel,\sigma)
= \\
& {\rm Tr}_C
\left[
\Gamma_{LL}(E;k_\parallel)\
G^r_{LR}(E;k_\parallel)\
\Gamma_{RR}(E;k_\parallel)\
G^a_{RL}(E;k_\parallel)
\right] \ ,
\end{split}
\label{eq:ToE}
\end{equation}
where (for simplifying the notation, explicit dependence of spin $\sigma$ 
and bias $V$ has been dropped)
the trace is taken over the basis set of the central $C$ region.
The GF
$G^{r/a}$ are the retarded/advanced NEGF of the $C$ region connected
to the leads, i.e. $G^{r/a}= [ g^{r/a} - \Sigma^{r/a}_{LL} - \Sigma^{r/a}_{RR} ]^{-1}$,
where $\Sigma^{r/a}_{LL}$ and $\Sigma^{r/a}_{RR}$ are the corresponding $L$ and $R$
lead self-energies respectively ($g^{r/a}$ is the GF of the disconnected
$C$ region). 
The quantities $\Gamma_{LL/RR}$ are the imaginary part of the lead self-energies,
$\Gamma_{LL/RR} = {\rm i} (\Sigma^{r} - \Sigma^{a})_{LL/RR}$.
Finally, $G_{LR}$ are the NEGF matrix elements connecting the most-left PL
of the $C$ region to the most-right PL of region $C$.

In order to calculate the supercurrent in the junctions, we actually need the full scattering matrix $S_N$ of the
central region.  The $S_N$ matrix is built from the transmission (reflection) coefficients between the $L$ and $R$ and
not from the transmission (reflection) probability Eq.~(\ref{eq:ToE}).  
{This marks one essential difference
between describing superconducting and normal transport: $S_N$ has additional information not needed for transmission in the
  normal state.}

Hence, we need to apply a transformation to Eq.(\ref{eq:ToE}) to be able to extract the transmission coefficients.
Instead of folding down the degrees of freedom of the $L$ and $R$ leads into a closed form for the lead self-energies
and calculating the transmission probability from a trace over the degrees of freedom of the central $C$ region, we have
to unfold these degrees of freedom and calculate the transmission probability from a trace over the these degrees of
freedom. The latter can form propagating waves in the $L$ and $R$ leads which are linked by transmission and reflection
coefficients as in the original picture of Landauer-like scattering \cite{Fisher:1981}.

Therefore we first need to determine the eigenmodes of propagation in the $L$ and $R$ leads, and then transform the lead
self-energies into the eigenmode basis of these propagating states.

\subsection{Eigenmodes of propagation in the leads}
\label{GFs}

\subsubsection{Bulk GF}
\label{bulkGF}

Deep inside the leads, 
one can calculate the eigenmodes of propagation by
solving a nearest-neighbor (in terms of PLs) tight-binding-like 
equation for the bulk GF $g_{p,p'}$
\begin{equation}
-S_{0,-1} g_{-1,0} + (P-S)_{0,0} g_{0,0} - S_{0,1} g_{1,0} = 1 \ .
\end{equation}
In the LMTO language \cite{Pashov:2020}, the quantity $(P-S)_{0,0}$ plays the role of a local 
energy-dependent Hamiltonian in the PL 
(with index $p=0$), where $P$ are the so-called potential functions 
and the structure constant $S_{p,p'}$ couples only adjacent PLs ($p-p' = \pm 1$).
The bulk is translationally invariant in the direction perpendicular to the PLs; hence
$S_{0,-1} = S_{1,0}$ and $g_{-1,0} = g_{0,1}$.

One solves the equation
\begin{equation}
-S_{1,0} g_{0,1} + (P-S)_{0,0} g_{0,0} - S_{0,1} g_{1,0} = 1 
\label{eq:TBgf}
\end{equation}
by expanding the wavefunction coefficients as a solution $\bm{\alpha}$ of a quadratic equation \cite{Chen:1989}.
This quadratic equation can be recast into a generalized eigenvalue problem $\bm{A} \bm{x} = \lambda \bm{B} \bm{x}$ 
by introducing
a new vector $\bm{\beta}=\lambda\bm{\alpha}$
and working in a enlarged (doubled) vector space \cite{Chen:1989,Fujimoto:2003}.
The generalized eigenvalue problem is written as:
\begin{equation}
\left[
\begin{array}{cc}
-{S}_{1,0} & (P-S)_{0,0} \\
{0} & {S}_{0,1} 
\end{array}
\right]
\left[
\begin{array}{c}
\bm{\alpha} \\
\bm{\beta}
\end{array}
\right]
=
\lambda 
\left[
\begin{array}{cc}
{0} & {S}_{0,1} \\
{S}_{0,1} & {0}
\end{array}
\right]
\left[
\begin{array}{c}
\bm{\alpha} \\
\bm{\beta}
\end{array}
\right] \ ,
\label{eq:GenEig_P-S_notation}
\end{equation}
and can be solved from a set of two independent equations:
\begin{equation}
-S_{1,0} P r^{-1} P^{-1} + (P-S)_{0,0} - S_{0,1} P r P^{-1} = 0 
\label{eq:PrP}
\end{equation}
and
\begin{equation}
-S_{1,0} Q x^{-1} Q^{-1} + (P-S)_{0,0} - S_{0,1} Q x Q^{-1} = 0 \ .
\label{eq:QxQ}
\end{equation}
Note that all matrices $M \equiv S,g, Q, P$ are dependent on variables
$M \equiv M(E;k_\parallel,\sigma)$. The above equations need to be solved
for each energy $E$, each $k_\parallel$ and each spin $\sigma$.

The eigenvalue vectors $r$ and $x$ characterize the propagating (or decaying)
modes in the bulk. Their meaning becomes clear when written as Bloch-like factors $e^{\pm i k_za}$,
where $a$ is the characteristic width of the PL and $k_z$ is the (energy dependent) wave number
normal to the interface. The columns of the matrices $P$ and $Q$ are the corresponding eigenvectors.
For the propagating modes, the wave numbers $k_z(E;k_\parallel,\sigma)$ are real numbers; 
$k_z(E)$ contains a non-zero imaginary part for the decaying modes.
We choose the following convention:
$\vert r_i \vert \le 1$ (propagating and decaying modes towards the right)
and
$\vert x_i \vert \ge 1$ (propagating and decaying modes towards the left).

\subsubsection{Reformulating the generalized eigenvalue problem}
\label{geneig}

One can rewrite the central equations of the previous section 
in term of the wavefunctions $\phi_p$ of
the PL $p$ deep inside the leads:
\begin{equation}
-S_{1,0} \phi_{p-1} +  g_b^{-1}\phi_{p} + S_{0,1} \phi_{p+1} = 0
\label{eq:diffEq_notation}
\end{equation}
as the system is translation-invariant in the bulk of the leads.
For notation convenience, we use $g_b^{-1}=(P-S)_{0,0}$.

By introducing the ratio matrix ${R}_p^{\leftarrow}=\phi_{p-1}\phi_{p}^{-1}$ and
by manipulation of the recursion relation 
\begin{equation}
\phi_{p} = g_b S_{1,0} \phi_{p-1} + g_b S_{0,1} \phi_{+1} \ ,
\label{eq:our_diffEq_2}
\end{equation}
one ends up with the following expression \cite{Fujimoto:2003} for the ratio matrix ${R}_{p+1}^{\leftarrow}$:
\begin{equation}
{R}_{p+1}^{\leftarrow} = 
\left({1} +  g_b S_{1,0} [{1} - {R}_p^{\leftarrow} g_b S_{1,0}]^{-1} {R}_p^{\leftarrow}\right) g_b S_{0,1} \ .
\label{eq:our_recur_3d}
\end{equation}
As the bulk of the lead is translation-invariant, we can use the generalized Bloch condition $\phi_{p+1} = \lambda \phi_p$ 
between adjacent PLs (note that this is reminiscent of the relation $\bm{\beta}=\lambda\bm{\alpha}$
used in the previous section).
The ratio matrix becomes ${R}_p^{\leftarrow}=\phi_{p-1}\phi_{p}^{-1}=\lambda^{-1}$, and Eq.~(\ref{eq:our_recur_3d})
is another formulation of the quadratic equation \cite{Chen:1989} discussed previously.

However, we have found this expression can be used as an iterative scheme, i.e.
\begin{equation}
\lambda_{i+1}^{-1} = 
\left({1} +  g_b S_{1,0} [\lambda_i - g_b S_{1,0}]^{-1} \right) g_b S_{0,1}
\end{equation}
to improve (at will) the accuracy (precision) of the eigenvalues $\lambda\equiv(r,x)$.
This is particularly crucial in the cases of degenerate modes that might occur at particular
$E$ and $k_\parallel$.

\subsubsection{The case of singular matrices}
\label{singular}

We have also implemented the possibilities of dealing with singular matrices that may
occur in the generalized eigenvalue problem $\bm{A} \bm{x} = \lambda \bm{B} \bm{x}$ 
which is not easily solvable when one, or both, $\bm{A}, \bm{B}$ are singular.
An eigenvalue shift procedure can be used to solve the linear generalized eigenvalue problem 
when both matrices are singular and when the full eigensystem is required \cite{Jennings:1978}.

For that one adds the term $-\alpha \bm{B} \bm{x}$ on both sides of $\bm{A} \bm{x} = \lambda \bm{B} \bm{x}$.
If $\bar{\bm{A}}=\bm{A} - \alpha \bm{B} $ is not singular, one can calculate
$\bar{\bm{M}}=\bar{\bm{A}}^{-1} \bm{B}$ and solve the conventional eigenvalue problem:
\begin{equation}
\bar{\bm{M}} \bm{x} = \frac{1}{\lambda - \alpha} \bm{x}.
\label{gen_eig_prb_M}
\end{equation}
A typical eigenvalue $\gamma$ of $\bar{\bm{M}}$ must be related to one of the $\lambda$ values
according to $\gamma = 1 / \mu$ where $\mu = \lambda - \alpha$, with the corresponding eigenvector
unchanged.

\subsection{Surface Green's functions}
\label{surfGF}

We can now build the surface GF of the leads which enters into the
definition of the lead self-energies $\Sigma^{r/a}_{RR}$.
The $L$ and $R$ surface GFs are obtained from only one part 
of the eigenmodes, i.e. from the propagating modes and the modes that
decay inside the bulk of the corresponding lead \cite{Chen:1989}.

The surface GF of the $R$ lead is obtained from
\begin{equation}
g_{RR} = \left[ g_{b}^{-1} - S_{0R} P r P^{-1} \right]^{-1}
\label{eq:gRR}
\end{equation}
and the $L$ lead surface GF from
\begin{equation}
g_{LL} = \left[ g_{b}^{-1} - S_{0L} Q x^{-1} Q^{-1} \right]^{-1} \ .
\label{eq:gLL}
\end{equation}
We have used an explicit notation for the leads' structure constant, $S_{0R}$ and
$S_{0L}$, as the two $L$ and $R$ leads need not be identical.

In comparison to the bulk case in Sec.~\ref{bulkGF}, we have
$S_{0R} = S_{0,1}$ and $S_{R0} = S_{1,0}$ for the $R$ region, 
and  
$S_{0L} = S_{1,0}$ and $S_{L0} = S_{0,1}$ for the $L$ region.
Note the different ordering of the subscript of the bulk structure constant 
for the $L$ and $R$ regions: 
the $R$ ($L$) surface GF is built from ``propagating''
towards two different $R$ ($L$) directions.
 
By introducing Eq.~(\ref{eq:PrP}) into Eq.~(\ref{eq:gRR}), and 
Eq.~(\ref{eq:QxQ}) into Eq.~(\ref{eq:gLL}),
we find that
the matrices $P,P^{-1}$ ($Q,Q^{-1}$) are the transformations that 
diagonalize $g_{RR} S_{R0}$ ($g_{LL} S_{L0}$) respectively.
Indeed, we have
\begin{equation}
g_{RR} = \left[ S_{R0} P r^{-1} P^{-1} \right]^{-1} \ ,
\label{eq:gRR_bis}
\end{equation}
or equivalently
\begin{equation}
g_{RR} S_{R0} = P r P^{-1} \ ,
\label{eq:gRR_ter}
\end{equation}
and 
\begin{equation}
g_{LL} S_{L0} = Q x^{-1} Q^{-1}  \ .
\label{eq:gLL_ter}
\end{equation}

The matrices $P,P^{-1}$ ($Q,Q^{-1}$) perform the change of basis set, from the original basis
to the basis of the propagation eigenmodes.

\subsection{Transmission coefficients}
\label{MW}

We now proceed with the transformation of the quantities $\Gamma_{\alpha\alpha}$ into
the basis of the propagation eigenmodes.
The quantity $\Gamma_{\alpha\alpha}$ is related to the imaginary part of the lead 
self-energy $\Sigma_{\alpha\alpha}$ ($\alpha{=}L,R$)
\begin{equation}
\Gamma_{\alpha\alpha} = S_{0\alpha}\ 
{\rm i} \left[ g^r_{\alpha\alpha}- g^a_{\alpha\alpha} \right]
S_{\alpha 0}  \ .
\label{eq:Gamma}
\end{equation}
Using Eq.~(\ref{eq:gLL_ter}) and the relation
$S_{0L} g^a_{LL} = (g^r_{LL} S_{L0})^\dag = (Q^{-1})^\dag (x^{-1})^\dag Q^\dag$,
we find that
\begin{equation}
\Gamma_{LL} = 
{\rm i}
\left(
S_{0L} Q x^{-1} Q^{-1} - (Q^{-1})^\dag (x^{-1})^\dag Q^\dag S_{L0} 
\right) \ .
\label{eq:GammaL_1}
\end{equation}
Introducing the identity $1=Q Q^{-1}$ ($1 = (Q^{-1})^\dag Q^\dag$)
to the right (left) side of the second (first) term in Eq.~(\ref{eq:GammaL_1}),
one ends up with:
\begin{equation}
\Gamma_{LL} = (Q^{-1})^\dag v_L Q^{-1}
\label{eq:GammaL_2}
\end{equation}
where
\begin{equation}
v_L = {\rm i}\left[
Q^\dag S_{0L} Q x^{-1}  - \left( Q^\dag S_{0L} Q x^{-1} \right)^\dag
\right] \ .
\label{eq:vL}
\end{equation}
Proceeding similarly for $\Gamma_{RR}$, we get
\begin{equation}
\Gamma_{RR} = (P^{-1})^\dag v_R P^{-1}
\label{eq:GammaR_2}
\end{equation}
where
\begin{equation}
v_R = {\rm i}\left[
P^\dag S_{0R} P r  - \left( P^\dag S_{0R} P r \right)^\dag
\right]
\label{eq:vR}
\end{equation}
It is crucial to note that the matrices $v_L$ and $v_R$ correspond to
the expectation values of the current operator calculated in the basis set 
of the eigenmodes \cite{Wimmer:2008}.

For non-degenerate modes, the diagonal elements $v_{L,n}$ ($v_{R,m}$) correspond 
to the group velocity $\partial E/\partial k$
of the propagating mode $n$ ($m$) in the $L$ ($R$) region \cite{Fujimoto:2003,Wimmer:2008}.
For decaying modes the diagonal elements are simply zero.
In the case of degeneracy, the velocity matrices $v_L$ and $v_R$
are block diagonal. Then we need to apply a further transformation to
get diagonal matrices $v_{L,R} \rightarrow v^D_{L,R}$
\cite{Wimmer:2008}.

Once the velocity matrices $v_L$ and $v_R$ are diagonal, we can
write
$v^D_{L,n}= \text{sgn}(v^D_{L,n}) \vert v^D_{L,n} \vert^{1/2} \vert v^D_{L,n} \vert^{1/2}$
and $v^D_{R,m}= \text{sgn}(v^D_{R,m}) \vert v^D_{R,m} \vert^{1/2} \vert v^D_{R,m} \vert^{1/2}$,
and transform Eq.~(\ref{eq:ToE}) as follows:

\begin{equation}
\begin{split}
& {\rm Tr}_C
\left[
\Gamma_{LL}\
G^r_{LR}\
\Gamma_{RR}\
G^a_{RL}
\right] \\
& =
{\rm Tr}
\left[
(Q^{-1})^\dag v^D_L Q^{-1}
 g^r_{LR}
(P^{-1})^\dag v^D_R P^{-1}
 g^a_{LR}
\right] \\ 
& = 
{\rm Tr}_{L+R}
\left[\ t_{LR} \ t_{LR}^\dag\ \right] \ .
\end{split}
\label{eq:ToE_bis}
\end{equation}
The transmission probability is now expressed in terms of transmission coefficients 
$t_{n,m}$ linking the propagating modes $n$ of the $L$ lead to the propagating
modes $m$ of the $R$ lead.
The transmission coefficients are given by a generalized Fisher-Lee expression \cite{Fisher:1981}:
\begin{equation}
\begin{split}
& t_{LR} \equiv \\ 
& t_{n,m}(E;k_\parallel,\sigma) =
{\rm i}
\vert v^D_{L,n} \vert^{1/2} \left[ {Q^{-1}\ g^r_{LR}\ (P^{-1})^\dag} \right]_{n,m} \vert v^D_{R,m} \vert^{1/2}
\end{split}
\label{eq:t_FL}
\end{equation}

It is important to note that the original Meir and Wingreen expression \cite{Meir:1992}
involves a trace over the degrees of freedom in the $C$ region (first line in Eq.~(\ref{eq:ToE_bis}) ),
while the Fisher and Lee expression \cite{Fisher:1981} involves a trace over the propagating modes
in the $L$ and $R$ leads (last line in Eq.~(\ref{eq:ToE_bis}) ).

\subsection{Reflection coefficients}
\label{rcoef}

In analogy to the transmission probability, we can also define reflection probabilities
in the same lead:
\begin{equation}
\begin{split}
R_L(E) &= {\rm Tr}_C
\left[
\Gamma_{LL}\
G^r_{LL}\
\Gamma_{LL}\
G^a_{LL}
\right] \\
R_R(E) &= {\rm Tr}_C
\left[
\Gamma_{RR}\
G^r_{RR}\
\Gamma_{RR}\
G^a_{RR}
\right] 
\end{split}
\label{eq:RoE}
\end{equation}

Following the derivations given in the previous section we find (for the $L$ region):
\begin{equation}
\begin{split}
& {\rm Tr}_C
\left[
\Gamma_{LL}\
G^r_{LL}\
\Gamma_{LL}\
G^a_{LL}
\right] \\
& =
{\rm Tr}
\left[
(Q^{-1})^\dag v_L Q^{-1}
 g^r_{LL}
(Q^{-1})^\dag v_L Q^{-1}
 g^a_{LL}
\right] \\
& = 
{\rm Tr}_{L}
\left[ r_{LL} (r_{LL})^\dag
\right] 
\end{split}
\label{eq:RoE_bis}
\end{equation}
The reflection coefficients $r_{LL}=r_{n,n'}$ are now expressed in the basis of the propagating
modes $n$ and $n'$ of the $L$ lead.

Similarly, one can find the reflection probability 
$R_R(E) = {\rm Tr}_{R} \left[ r_{RR} (r_{RR})^\dag \right]$
from the reflection coefficients $r_{RR}=r_{m,m'}$ 
expressed in the basis of the propagating
modes $m$ and $m'$ of the $R$ lead.

One should note that the reflection probability, defined in Eq.~(\ref{eq:RoE}),
contains the contributions of both the $L$ ($R$) incoming wave(s) and the reflected 
waves in the $L$ ($R$) region.
In order to obtain the correct reflection coefficients (and proper flux conservation),
one needs to suppress the contribution of the incoming wave in the $n$-th $L$ channel 
($m$-th $R$ channel) in the $L$ ($R$) lead respectively.

Hence the expressions of the reflection coefficients are as follows:
\begin{equation}
\begin{split}
& r_{LL}=
r_{n,n'}(E;k_\parallel,\sigma) = \\
& {\rm i}
\vert v^D_{L,n} \vert^{1/2}\ \left[ Q^{-1}\ g^r_{LL}\ (Q^{-1})^\dag \right]_{n,n'} \vert v^D_{L,n'} \vert^{1/2}
 - \delta_{nn'} 
\end{split}
\label{eq:rcoefL}
\end{equation}
and 
\begin{equation}
\begin{split}
& r_{RR}=
r_{m,m'}(E;k_\parallel,\sigma) = \\
& {\rm i}
\vert v^D_{R,m} \vert^{1/2}\ \left[ P^{-1}\ g^r_{RR}\ (P^{-1})^\dag \right]_{m,m'} \vert v^D_{R,m'} \vert^{1/2}
 - \delta_{mm'}
\end{split}
\label{eq:rcoefR}
\end{equation}

\section{Full scattering matrix and supercurrent}
\label{sec:fullSmatrix}

The full scattering matrix $S_N$ is built from the reflection coefficients $r_{LL}$ and $r_{LL}$, and
from the transmission coefficients $t_{LR}$ and $t_{RL}$ ($t_{RL}$ is the transpose of $t_{LR}$).
All quantities are explicitly dependent of $(E;k_\parallel,\sigma)$.

For the calculations of the supercurrent, we construct the 
normal state scattering matrix $S_N$ as follows:

\begin{eqnarray}
S_N(E;k_\parallel)=
\left(
\begin{array}{cc}
\left[
\begin{array}{cc}
r_{LL}(\uparrow) & 0 \\
0 & r_{LL}(\downarrow)
\end{array}
\right] & 
\left[
\begin{array}{cc}
t_{LR}(\uparrow) & 0 \\
0 & t_{LR}(\downarrow)
\end{array}
\right] \\
 \\
\left[
\begin{array}{cc}
t_{RL}(\uparrow) & 0 \\
0 & t_{RL}(\downarrow)
\end{array}
\right] & 
\left[
\begin{array}{cc}
r_{RR}(\uparrow) & 0 \\
0 & r_{RR}(\downarrow)
\end{array}
\right] \\
\end{array}
\right) \nonumber \\
. 
\label{eq:fullSmatrix}
\end{eqnarray}
The normal state scattering matrix $S_N(E_F+\varepsilon;k_\parallel)$ characterizes electron (particle)
transport for a positive energy $\varepsilon$ above the Fermi level $E_F$.
The transport of hole (antiparticle) is given by the time-reserved symmetric $S_N^*(E_F-\varepsilon;k_\parallel)$
scattering matrix for (negative) energy $-\varepsilon$ below $E_F$.

As mentioned above, the dc Josephson current in the junction is obtained from the Andreev bound states formed
in the junction.
The spectrum of the Andreev bound states can be calculated
from a scattering matrix formalism \cite{Beenakker:1992,Beenakker:2004}
In such an approach, the spatial separation of Andreev and
normal scattering is the key simplification which allows one
to relate the Josephson current directly to the normal-state
scattering matrix of the junction.

We have successfully applied such an approach in a recent paper \cite{Ness:2022}
to the study of the supercurrent decay and oscillation in magnetic Josephson
junctions made of Ni layers connected to two Nb leads.

\section{Application}
\label{sec:bulkco}

We now provide an illustrative example of our eigenmode approach
applied to a simple case. We consider that all PLs in the $L$-$C$-$R$ junction
are identical, i.e. we study the bulk (equilibrium) transport properties of a
``perfect'' crystal. In this case, the transmission probability (at a given
energy $E$ and for a given $k_\parallel$ point) is simply given by the number of ``real''
bands crossing that energy $E$ at that $k_\parallel$ point, each ``real'' band corresponding
to a propagating mode. 
We compare the
results obtained from the two Meir and Wingreen \cite{Meir:1992} and Fisher and Lee \cite{Fisher:1981}
expressions.

\begin{figure}
	\centering
	\includegraphics[scale=0.35]{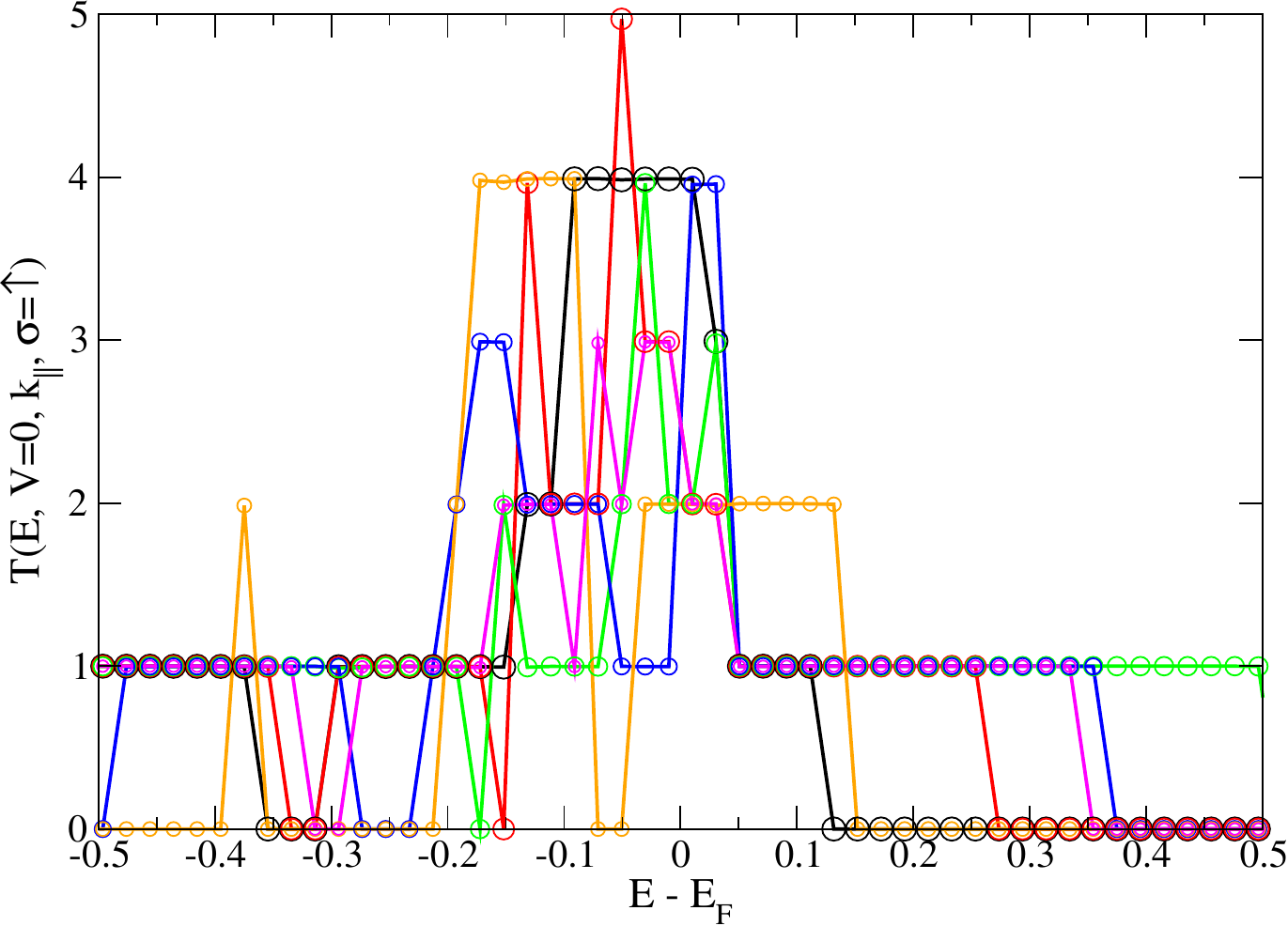}
	\caption{Transmission probability $T(E,V=0;k_\parallel,\sigma=\uparrow)$ for an energy window around
	the equilibrium Fermi level $E_F=0$ and for different $k_\parallel$ points (corresponding to different
	colors). The solid lines are from
	the Meir and Wingreen expression. The open circles correspond to the Fisher and Lee formula for
	which only 5 propagating modes are taken into account.}
	\label{fig:Cobulk}
\end{figure}

Figure \ref{fig:Cobulk} shows the transmission probability $T(E,V;k_\parallel,\sigma)$ calculated
at equilibrium $V{=}0$ and for spin $\uparrow$ of bulk Co.
The unique PL of bulk Co is made of 2 atoms of Co with 9 (\emph{spd}) orbitals, i.e. the size
of the matrices corresponding to that PL is 18$\times$18 per spin.
The calculations performed from the transmission coefficients with only 5 propagating modes provide
the very same results as the transmission probability obtained from Eq.(\ref{eq:ToE}), as expected.
Note that the maximum transmission probability is 5, which is indeed the number of propagating modes
we found.

We conclude the present paper by the following note:
the size ($N\times N$) of the velocity matrices $v_\alpha$ is given 
by the size of the matrices for the structure constant $S_{0\alpha}$ and 
for $g_b^{-1}$ in the identical PLs of the $\alpha=L,R$ leads (this
is for each spin and for each value of $k_\parallel$).
We have found that the number of propagating modes (with non zero value of 
the velocity) is much smaller than $N$. 
This reduces considerably the size of the transmission (reflection) coefficient 
matrices and hence improve the computational performances.

For bulk Co, we have seen that $N=18$ and only 5 propagating modes are present. 
For other examples shown in \cite{Ness:2022}, we get similar trends.
In the case of Nb(110)/Ni(111)/Nb(110) junctions, each PL of
the $\alpha=L,R$ leads contains 10 atoms of Nb with 9 (\emph{spd}) orbitals,
hence $N=90$. However, there is only a maximum of 25 propagating modes in
the corresponding Nb leads.
For Nb(110)/Ni(110)/Nb(110) junctions, there are 2 atoms of Nb in each PL,
i.e. $N=18$, and only 7 propagating modes in the leads.
For Nb(110)/Fe(111)/Nb(110), there are 3 atoms of Nb in each PL, i.e. $N=27$,
and only 7 propagating modes in the leads \cite{Ness:unpub}.

\begin{acknowledgements}

HN and MvS acknowledge financial support from Microsoft Station Q
via a sponsor agreement between KCL and Microsoft Research.
In the late stages of this work MvS was supported by the U.S. Department of Energy, 
Ofﬁce of Science, Basic Energy Sciences under Award \# FWP ERW7906.
The authors acknowledge fruitful discussions and collaboration with
Roman Lutchyn, Ivan Sadovskyy and Andrey Antipov leading to Ref.~[\onlinecite{Ness:2022}].

\end{acknowledgements}

\newpage

\end{document}